\documentclass[preprint]{aastex}

\def\kms{{\rm km\ s^{-1}}}
\def\yr{{\rm yr}}
\def\AU{{\rm AU}}
\def\cm{{\rm cm}}

\shortauthors{Ostriker et al}
\shorttitle{Jet-driven outflow shells}

\begin{document}

\title{A ballistic bow shock model for jet-driven protostellar outflow shells}

\author{Eve C. Ostriker, Chin-Fei Lee, James M. Stone, 
and Lee G. Mundy}
\affil{Department of Astronomy, University of Maryland \\
College Park, MD 20742-2421}
\email{ostriker@astro.umd.edu, chinfei@astro.umd.edu,
  jstone@astro.umd.edu, lgm@astro.umd.edu}

\begin{abstract}
  We analyze the dynamics of the shell produced when a bow shock from
  a collimated jet propagates into the surrounding medium.  Under
  interstellar conditions, the shock is radiative, and a ballistic
  approximation for the shell flow is appropriate beyond the working
  surface where the jet impacts its surroundings.  The solution is
  then determined by the ambient and jet densities and velocities and
  by the momentum impulse applied in the working surface.  Using
  estimates for these impulses (confirmed by separate numerical
  simulations), we obtain solutions for the shell structure, and for
  the range of velocities in the shell at any point.  We provide
  predictions for the position-velocity and mass-velocity relations
  expected for plane-of-sky bow-shock shells, and for the bulk shell
  properties.  In a companion paper, we show that these analytic
  solutions are in excellent agreement with the results of direct
  numerical simulations.  We argue that classical molecular (CO)
  outflows cannot be purely jet-driven, because the bow-shock shell
  solutions are much too elongated compared with observations.
  Finally, we suggest that the ``spur'' structures seen in
  position-velocity diagrams of observed molecular outflows are the
  manifestation of internal bow shocks which may be fit with our simple
 dynamical models.
\end{abstract}


\section{Introduction}

Bipolar molecular outflows appear to be an inevitable byproduct of
low-mass star formation: essentially every pre-main sequence star that
is still surrounded by substantial molecular material shows signs of
outflow in the molecular line channel maps.  Although there is
considerable variety in the structure and kinematics of these outflow
lobes (including a great deal of irregular morphology) many of these
lobes take the shape of hollow shells (e.g. \cite{Moriarty1987}).  
In cases where optical or
infrared Herbig-Haro jets are also seen, the molecular shells are
approximately centered on the axes defined by these jets.  Recent
reviews of the molecular outflow phenomenon are found, e.g., in
\cite{Bachiller1999} and \cite{Richer2000}.

Two basic models have been advanced to explain how molecular shells
are driven from young stellar objects (YSOs).  In the first model, the
outflow is produced by a wide-angle wind directly sweeping up
surrounding ambient material (see discussion and original references
in \cite{Shu2000}). In the second model, the outflow represents an
expanding bow shock in the ambient medium produced by the impact of a
narrow, dense jet (see e.g. the semianalytic models of
\cite{Masson1993, Raga1993}, and the simulations by \cite{Chernin1994,
  Smith1997, Suttner1997, Downes1999}).  In the wide-angle wind
scenario, outflows are a purely momentum-driven phenomenon; the
thermal pressure of the gas is taken to be negligible.  In the
jet-driven scenario, on the other hand, the transverse (i.e.
perpendicular to the jet axis) expansion of a bow shock to create an
outflow shell depends crucially on the action of pressure in the head
of the jet.  The two models thus differ both in their assumptions of
the nature of the primary wind, and in the physical processes involved
in the ambient medium interaction that produces the outflow.

In following a fully deductive theoretical approach, one would first
determine the nature of the primary wind that is expected to form, and
then analyze how it acts on its surroundings to produce outflows.
However, a more inductive approach -- in which clues to the nature of
the primary wind may be gleaned from analyzing the ``secondary''
outflow kinematics -- can often aid progress in circumstances, such
as the present one, where the first-principles approach involves major
theoretical challenges.  In particular, although there is significant
theoretical consensus that the primary winds from low mass YSOs are 
magnetocentrifugally driven, there remain unresolved questions about the
relative importance of winds driven from the interaction region 
between a stellar magnetosphere and the disk (x-winds; \cite{Shu2000} and
references therein), and winds arising from a larger range of 
radii in the disk (disk winds; \cite{Konigl2000} and references therein).  

In fact, neither the x-wind nor the disk-wind model would produce the
sort of spatially-isolated ``pure jet'' generally taken as the input to a
jet-driven outflow calculations, because such a jet would be highly
magnetically overpressured relative to the ambient ISM at large
distance from the source \citep{Ostriker1997,Kim2000}.  Thus, at least
the surface layers, and possibly the entirety, of either sort of wind
would expand to produce a gradient of $B_\phi^2$ matching the low
pressure of the ISM on the outside and the high pressure of the wind
core on the inside.  \cite{Shu1995} show that a radially-expanding,
magnetically force-free ($B_\phi\propto R^{-1}$) wind with density
stratified as $\propto R^{-2}$ is a self-consistent solution to the
asymptotic state of x-winds; similar ``fully-expanded'' magnetically
force-free solutions may be found for the asymptotic state of disk
winds \citep{Ostriker1997,Ostriker1998,Matzner1999}.  
Another outcome potentially
could involve only partial expansion of the wind, leaving a highly
overdense core and strongly stratified surroundings.  Whether a
wind attains full or only partial expansion may depend on its
boundary conditions and stability properties (e.g. \cite{Kim2000}),
with no comprehensive theoretical predictions available at present.

Thus, although there is not yet a complete theoretical catalog for the
range of structure possible for primary winds,
first-principles theoretical considerations do suggest that completely
isolated jets would not occur in general.  Instead, the primary wind
in the ``jet-driven'' outflow scenario is more properly thought of as
a more extreme (in its degree of central concentration) version of the
radially flowing primary wind invoked in the wide-angle wind model.
``Jet-driven'' outflows, then, would arise if the circumstances were
such as to produce a primary wind with a very dense core (the optical
``jet'') surrounded by an (optically unseen) envelope in which the
density (and magnetic pressure) falls off very steeply (e.g. faster
than an $R^{-2}$ power law).  With this modified view of the nature of
primary wind, any model should contain a wide-angle wind at some
level.  Two classes of outflow-driving scenarios might then contrast
``wind-swept shell'' models vs. ``bow shock shell'' models based on
the relative importance of transverse bulk flow momentum in the
primary wind (i.e. $\rho v_R^2$ ram pressure), {\it vs.} transverse
thermal pressure gradients created in the shock, in producing the
expansion of the outflow away from the central axis.

The wind-swept shell formalism was first presented in the x-wind
context by \cite{Shu1991}, and later updated to reflect the strong
polar- and equatorial-  stratification subsequently found to arise,
respectively, in magnetic wind and protostellar core solutions
\citep{Li1996}.  
\cite{Matzner1999} recently extended the wind-swept shell 
analysis to more general classes of
primary winds.  In all of these models, 
the primary wind is assumed
to be radially outflowing, and the interaction with the ambient medium
is treated as a local mass- and momentum- conserving process.  
With these idealizations, together with the assumption that the 
radial variation of the density 
profile in the ambient core follows an $r^{-2}$ power law,
one may obtain analytic solutions for the dynamics
of wide-angle-wind driven outflow shells in which the shell velocity is 
steady but depends on the polar angle.  These
models have proven very successful at explaining the shapes and
kinematics of many -- but not all -- of the features seen in the
outflows from young stars 
\citep{Li1996,Nagar1997,Ostriker1997,Matzner1999,Lee2000a}.  
In particular, several outflows
show evidence of ``convex spurs'' in position-velocity space, in which
transverse velocities of portions of the outflow increase with
distance from the source.  In configuration (position-position) space, 
the corresponding outflow shell closes toward the axis
with increasing distance, with the appearance of a bow shock
\citep{Lee2000a}.  

In this paper, our goal is to develop an analytic model 
for protostellar outflow shells  in the situation complementary to that 
addressed by \cite{Shu1991}:  namely, when the transverse radial forces
are dominated by pressure gradient forces near the 
the head of the jet (or, more properly, jet-like wind), 
rather than by ram pressure throughout the body of an extended wind.
To this end, we construct a simple dynamical model of jet-driven
bow shocks under interstellar (strongly-cooling) conditions.  The model
solutions depend only on four basic properties describing the jet and
ambient medium (the jet speed and radius, and the jet and ambient densities),
and on the cooling function for shocked gas.  We provide 
analytic solutions for the shell shape and velocity fields in this model, 
and use these results to develop solutions for 
two kinematic diagnostics often used in analyzing observations of 
outflows -- the position-velocity and mass-velocity relations.  We also 
provide expressions for ``bulk'' properties of jet-driven outflow shells
(lobe aspect ratio, total mass, momentum component ratios)
in terms of the system's physical parameters.
In the companion paper \citep{Lee2000b}, 
we compare the results from the 
analytic model of this paper with the results of direct numerical 
simulations of bow shocks driven by model protostellar jets.  There, we also 
compare the results of wind-swept shell ``snowplow'' 
models with the results of numerical simulations where the input wind 
has an extended (wide-angle) density and velocity distribution.

\section{Analysis: bow shock shape and shell velocities}

\begin{figure}
\center
\epsscale{0.6}
\rotatebox{-90}{\plotone{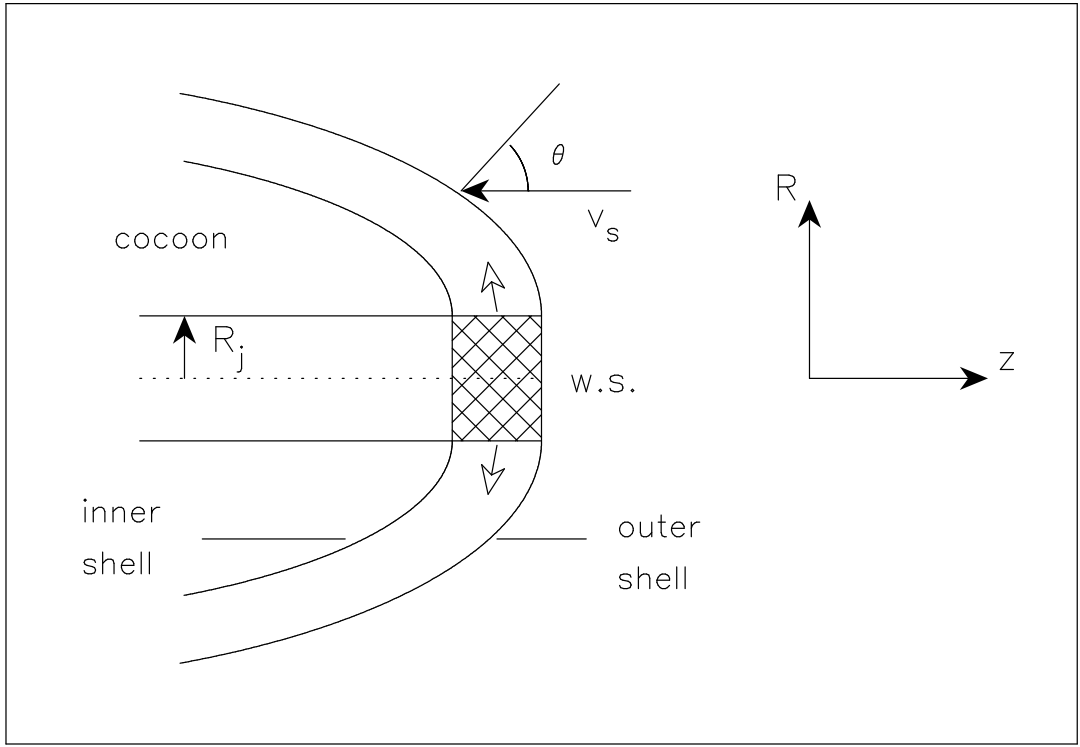}}
\caption{A schematic diagram of the model 
viewed from the bow shock frame moving at shock velocity $v_s$,
in a cylindrical coordinate system. Here ``w.s.'' is the
working surface around the hot shocked material at the jet head. 
In this frame, material flows out from the working surface into the shell, 
and ambient material flows onto the shell with velocity $-v_s \hat z$. 
The angle between the local normal to the shell and the $\hat z$ direction 
is $\theta$.
\label{fig:impulse}}
\end{figure}

The interaction of a supersonic jet
with its surroundings drives a bow shock into the ambient medium;
under uniform conditions with time-independent jet properties, the bow
shock preserves its shape and advances along the jet axis $\hat z$ at
a speed $v_s$.  Figure \ref{fig:impulse} shows a schematic diagram of
our model for shell formation, viewed from the bow shock frame
moving at velocity $v_s \hat z$ with respect to the observer's frame.  We
adopt a cylindrical coordinate system. In this diagram, the working
surface (``WS'') is the region where the jet itself collides and interacts 
with the ambient medium: hot, shocked jet material processed through a jet
shock (at the left of the WS) abuts hot, shocked ambient material
processed through the bow shock (at the right of the WS) at a contact 
discontinuity (in the center of the WS).  
Due to thermal pressure gradient forces perpendicular to the jet axis, 
material is expelled radially from the WS.  This flow expands away
from the jet, and drives a bow shock into the ambient medium.  The ambient
material, moving backward in the frame of the bow shock, collides with the
shell of transversely-flowing matter and deflects it rearward (in the 
$-\hat z$ direction).  The swept-up ambient material augments the 
flux of matter in the transversely-expanding flow.

The shape of the shell and simple kinematic diagnostics can be
derived by solving for the shell dynamics in the ballistic limit,
i.e. with pressure forces ignored subsequent to an initial impulse.
\footnote{This approximation is valid because the shock (away from the 
jet head) is sufficiently oblique that the post-shock pressure is 
is relatively low, and is further reduced by the strong cooling present 
under interstellar (compared to extragalactic) conditions.}
In the bow shock frame, the ambient material flows into
the shock with velocity $-v_s \hat z$. The increase in 
the mass flow $\dot M$ of shocked gas 
per unit length transverse to the jet axis $dR$ is
\begin{equation}\label{eq:jetM}
\frac{d\dot{M}}{dR}=2\pi R \rho v_s,
\end{equation}
where $\rho$ is the density of the ambient material.
Similarly, the increases per unit transverse length in the 
axial-direction and radial-direction flows of momentum 
($\dot{P}_z$, $\dot{P}_R$) are  
\begin{equation}
\label{eq:jetz}
\frac{d\dot{P}_z}{dR}=- 2\pi R \rho v_s^2
\end{equation}
and 
\begin{equation}
\label{eq:jetR}
\frac{d\dot{P}_R}{dR}=0.
\end{equation}

Integrating equations (\ref{eq:jetM}), (\ref{eq:jetz}) and (\ref{eq:jetR}),
we have
\begin{equation}
\label{eq:IjetM}
\dot{M}=\dot{M}_o+\pi \rho v_s (R^2-R_o^2)
\end{equation}
\begin{equation}
\label{eq:Ijetz}
\dot{P}_z=\dot{P}_{oz}- \pi \rho v_s^2 (R^2-R_o^2) 
\end{equation}
\begin{equation}
\label{eq:IjetR}
\dot{P}_R=\dot{P}_{oR}
\end{equation}
where $R_o$ is the radial width of the WS where the shell 
emerges.
$\dot{M}_o$, $\dot{P}_{oz}$ and
$\dot{P}_{oR}$ are
the initial mass, longitudinal momentum and transverse momentum flows 
input from the WS into the shell.

If a negligible portion of the shocked-gas 
mass and momentum are lost from the inner shell 
into the cocoon, then the total shell mass flow and momentum flow at a
distance $R$ are $\dot{M}$, $\dot{P}_{z}$ and
$\dot{P}_{R}$. In the absence of shell mass losses, the velocity field 
{\it direction} within 
the shell must be locally parallel to the shell surface (the bow shock),
although there may be gradients in the {\it magnitude} of the velocity 
across the shell thickness perpendicular to the shell surface.
Under this assumption, the ratio
$v_z/v_R$ is the same throughout the shell thickness at any $R$, and equal
to the slope $dz/dR$ of the shell's surface.  Since $v_z/v_R$ is constant
across the shell thickness, the ratio of the total momentum flows 
$\dot P_z =\int v_z (R,s)\, \rho\ 2\pi R ds$ and 
$\dot P_R =\int v_R (R,s)\, \rho\ 2\pi R ds= (v_R/v_z)\dot P_z $  also gives
the local slope of the shell surface (here the integral over $s$ denotes
summation across the shell thickness).
The shape of the shell can thus be derived by integrating 
\begin{equation}
\label{eq:jets}
\frac{dz}{dR}= \frac{\dot{P}_z}{\dot{P}_R}=
 \frac{\dot{P}_{oz} - \pi \rho v_s^2 (R^2 - R_o^2)}{\dot{P}_{oR}}.
\end{equation}
The locus of the bow shock/outflow shell is therefore given by
\begin{equation}
\label{eq:Ijets}
z= \frac{\dot{P}_{oz} (R-R_o)- \pi \rho v_s^2  (R^3/3 - R R_o^2 + 2R_o^3/3) }
{\dot{P}_{oR}}
\end{equation}
with $z=0$ at $R=R_o$.

With expression (\ref{eq:Ijets}) giving the locus of the outflow shell,
we may now turn to the distribution of velocities within the shell.
Consider first the limit in which the material added to the shell at
any point mixes instantaneously with all the material already flowing
along in the shell at that position.  If the ``new'' and ``existing''
momentum were to mix thoroughly, then there would be no velocity
gradients across the thickness of the shell.
The velocity of shell material in the shock frame 
would be equal to the {\it mean} value at any point, given by
\begin{equation}
\bar{v}_z=\frac{\dot{P}_z} {\dot{M}} = 
  \frac{\dot{P}_{oz} - \pi \rho v_s^2  (R^2 - R_o^2)}
       {\dot{M}_o + \pi \rho v_s (R^2 - R_o^2)}
\end{equation}
\begin{equation}
\bar{v}_R=\frac{\dot{P}_R} {\dot{M}} = 
  \frac{\dot{P}_{oR}}{\dot{M}_o + \pi \rho v_s  (R^2 - R_o^2)}
\end{equation}
In the observer's frame, $\bar{v}_z$ transforms to 
\begin{equation}
\bar{v}_z'=\bar{v}_z + v_s= 
  \frac{\dot{P}_{oz} +\dot{M}_o v_s}
       {\dot{M}_o + \pi \rho v_s  (R^2 - R_o^2)}
\end{equation}
In particular, the shock-frame shell velocities just outside the WS are 
$\bar v_R=\dot{P}_{oR}/\dot{M}_o$ and 
$\bar v_z=\dot{P}_{oz}/\dot{M}_o$, where the latter transforms to
$\bar v_z'=v_s + \dot{P}_{oz}/\dot{M}_o$ in the observer frame.

In the simulations presented by \cite{Lee2000b} in a companion paper, 
we find that 
the newly swept-up ambient material does {\it not } in fact 
fully mix with the material already in the shell. We are thus led to 
analyze the opposite limit from the above:  no mixing between ``new'' and 
``existing'' shell material.  
Consider first the outer surface layer of the shell consisting of the
material that has most recently been swept up by the advancing bow shock.
If the ambient material shocks strongly
upon colliding with the shell, the postshock velocity of the newly swept-up
material in the shock frame at any point 
is just the component of preshock velocity
parallel to shell surface. With 
$\theta\equiv-\textrm{arctan}(\frac{dz}{dR})$ 
the angle between the shock normal and $\hat{z}$
(see Fig. \ref{fig:impulse}), the velocity 
of the newly-shocked material is given by
\begin{equation}
{\bf u}=v_s \sin \theta(-\sin\theta \hat{z} +\cos\theta\hat{R})
\end{equation}
in the shock frame.
In the observer's frame, ${\bf u}$ transforms to
\begin{equation}
{\bf u}'= {\bf u} + v_s \hat{z} = v_s (\cos^2 \theta \hat{z}
+\sin\theta\cos\theta\hat{R})
\end{equation}
The shape of the shell $z(R)$ is determined by the total momentum flows only,
independent of whether momentum is mixed within the shell or not.
Therefore, substituting from  equation (\ref{eq:jets}) for $-\tan\theta$,
we find 
the velocity components of the newly swept-up material in the observer's 
frame are
\begin{equation}
u_z' = \frac{v_s}{1+\left(
 \frac{\dot{P}_{oz} - \pi \rho v_s^2 (R^2 - R_o^2)}{\dot{P}_{oR}}\right)^2}
\end{equation}
and
\begin{equation}\label{eq:uR}
u_R = \frac{-v_s \left(
 \frac{\dot{P}_{oz} - \pi \rho v_s^2 (R^2 - R_o^2)}{\dot{P}_{oR}}\right)}
{1+\left(\frac{\dot{P}_{oz} - \pi \rho v_s^2 (R^2 - R_o^2)}
{\dot{P}_{oR}}\right)^2}.
\end{equation}

In the limit of negligible mixing, the {\it magnitude} of the momentum 
$\bf p$ of any fluid element (and hence its speed) would remain unchanged as
it flows outward along the bow-shock shell, because the impact of new
material to the shell applies a force transverse to its surface 
(but not {\it along} it) as it shocks. With a force perpendicular to 
$\bf p$ (which is parallel to the shell surface), the
magnitude $|{\bf p}|$ would not change, but the direction would 
rotate always 
remaining parallel to the shell surface.   Thus, at the point $R$, 
the material that entered the shell from the WS  with 
speed $v_0\equiv(\dot P_{oz}^2+ \dot P_{oR}^2)^{1/2}/\dot M_o$ 
would  have shock-frame
component velocities 
$w_R(R;R_0)=v_0 \cos\theta$
and 
$w_z(R;R_0)=-v_0 \sin\theta$
at the position $R$, with the latter transforming to
$w_z'(R;R_0)=-v_0 \sin\theta + v_s $
in the observer's frame.  Similarly, 
the material that entered the shock at $R'$ 
with post-shock speed $v(R')=v_s \sin\theta'$ would
have components at position $R$ given by 
$w_R(R;R')=v_s \sin\theta' \cos\theta$
and $w_z(R;R')=-v_s \sin\theta' \sin\theta$, 
with the latter transforming to
$w_z'(R;R')=v_s(1-\sin\theta' \sin\theta)$
in the observer's frame; here $\theta'\equiv -\arctan (dz/dR)|_{R'}$.  
Since $\sin\theta'$ is a secularly increasing function of $R'$,  
$w_R(R;R')$ increases from $0$ to  $u_R(R)$ and
$w_z'(R;R')$ decreases from $v_s$ to $u_z'(R)$ as $R'$ increases from 
just outside $R_0$ to $R$, 
for fixed $R$.  That is, the more recently a fluid element has joined
the shell flow from the ambient medium, 
the larger its transverse and the smaller its longitudinal
velocity would be at any point $R$, in the absence of mixing.

To obtain explicit solutions, we now relate the shell input flows
$\dot{M}_o$, $\dot{P}_{oz}$ and $\dot{P}_{oR}$ to more basic
quantities characterizing the jet and ambient medium.  First, we note
that for a strong shock, the bow shock speed is related to the jet
speed $v_j$ and ratio $\eta\equiv \rho_j/\rho$ of jet- to ambient-
density by $v_s\approx v_j (1+\eta^{-1/2})^{-1}$.  The radius of the
WS can be approximated as $R_o \approx R_j$.  As shown in Figure
3 of \cite{Lee2000b}, only shocked ambient material
flows into the shell from the WS; therefore the mass flow is just
equal to the flow of ambient material into the WS, $\dot{M}_o\approx
\pi R_j^2 \rho v_s$ .  The input longitudinal momentum flow
$\dot{P}_{oz}\approx 0$, because in the shock frame, the ejected flow
from the WS has no preferred forward or backward direction.

Estimating the transverse momentum flow is slightly more subtle.  Just
inside the bow shock, the shocked ambient gas cools rapidly down to
$\sim 10^4$K, after which point the radiation slows because the
cooling curve drops precipitously as the gas recombines. 
The shocked flow from the ambient medium
interfaces (in a contact discontinuity) with the shocked jet gas at
the center of the WS; in a steady state and with negligible mixing,
all of the streamlines entering the WS from either side must bend away
from the axis and exit on their respective sides of the bow shock/jet
shock interface.  The pressure is highest closest to the axis, with
the transverse pressure gradient accelerating gas radially to eject it
from the sides of the WS.  

The ejection velocity from the WS must be
of order the sound speed $c_s$ at the temperature $10^4$K.  To see why
this is so, consider adiabatic flow with $\gamma=5/3$ starting from
low velocity $v$ and high pressure $P$, and accelerating by pressure
gradient forces until the ram pressure far exceeds the thermal
pressure.  By Bernoulli's theorem, which demands the constancy of
$(1/2)v^2 + [\gamma/(\gamma-1)]P/\rho$, the ejection speed would thus
be $\sim \sqrt{5} c_s$, where $c_s$ is the isothermal sound speed
$(kT/\mu)^{1/2}$ at the initial temperature of the flow. Here, $\mu$ is
the mean molecular mass, which we take as $1.3m_p$ for neutral gas.
\footnote{\cite{Falle1993} and \cite{Biro1994} have made 
related arguments, confirmed by simulations, for estimating the speed 
of material ejection from internal shocks in jet beams.}
To the extent that cooling reduces the
enthalpy term, and that the flow does not reach the maximum possible
speed, the outflow velocity from the working surface would be somewhat 
lower.  We therefore expect
a transverse momentum flow from the WS of order 
$\sqrt{5}\dot{M}_o c_s= 2.2\pi \rho R_j^2 v_s c_s$, where the value of
$c_s$ at $10^4$K is $8\rm km\, s^{-1}$.  
To test this estimate, we
have directly measured the transverse momentum flux in our simulations
with varying $v_j$ and $R_j$, and also fit the shell shapes and 
kinematics parameterized by the momentum flux (see below and \cite{Lee2000b}).
Writing $\dot{P}_{oR} \equiv \beta \pi
\rho R_j^2 v_s c_s$, we have found that $\beta=3.8-4.4 $ for jet radii
and velocities in the ranges $R_j=2.5-5 \times 10^{15}\ \cm$ and
$v_j=120-240\ \kms $.  These values of $\beta$ are slightly larger than
the above estimate;  the difference may be accounted for by the 
facts that (i) the pressure of the shocked jet gas adjoining the shocked 
ambient gas within the WS helps in part to accelerate the shell, 
and (ii) the ambient material just outside the WS also passes through
a relatively perpendicular (rather than oblique) shock, maintaining a
non-negligible pressure in the shell at radii slightly larger than 
$R_j$ and correspondingly raising the total transverse momentum flux 
delivered to the bow shock shell.  

With the above substitutions for $\dot{M}_o$, $\dot{P_{oR}}$, and
$\dot{P_{oz}}$, 
we find $\tan\theta=(v_s/\beta c_s)[(R/R_j)^2-1]$ so 
\begin{equation}
\label{eq:costheta}
\cos\theta=
\frac{\beta c_s R_j^2}
{\left[(\beta c_s R_j^2)^2 +
v_s^2(R^2 - R_j^2)^2\right]^{1/2}}
\end{equation}
and  
\begin{equation}
\label{eq:sintheta}
\sin\theta=\frac{v_s(R^2 - R_j^2)}
{\left[(\beta c_s R_j^2)^2 +
v_s^2(R^2 - R_j^2)^2 \right]^{1/2}}.
\end{equation}
The components of the mean shell velocity $\bar{\bf v}'$ and
newly-shocked velocity ${\bf u}'$ in the observer's frame are given by
\begin{equation}\label{eq:meanvelz}
\bar{v}_z'=\frac{R_j^2}{R^2}v_s,
\end{equation}
\begin{equation}\label{eq:meanvelR}
\bar{v}_R=\frac{R_j^2}{R^2}\beta c_s,
\end{equation}
\begin{equation}\label{eq:newvelz}
u_z' = \frac{(\beta c_s R_j^2)^2}{(\beta c_s R_j^2)^2 +v_s^2(R^2-R_j^2)^2} v_s,
\end{equation}
and
\begin{equation}\label{eq:newvelR}
u_R = \frac{\beta c_s v_s R_j^2(R^2-R_j^2)}
{(\beta c_s R_j^2)^2 +v_s^2(R^2-R_j^2)^2} v_s.
\end{equation}
The shape is given by
\begin{equation}\label{eq:shapesol}
z= - \left[{1\over 3}\left({R\over R_j}\right)^3 - {R\over R_j} + 
{2\over 3}\right]{v_s\over \beta c_s} R_j,
\end{equation}
which at large distance from the head of the jet approaches a cubic law
\footnote{A similar asymptotic $|z|\propto R^3$ law was previously obtained
by \cite{Wilkin1996} in his solution for the shell shape in the ``tail'' of
a stellar wind bow shock for a star moving with constant velocity through a
uniform medium.  This is consistent with expectations that the
specific geometry of the transverse momentum source is not important for the
far-field solution.},
$z\approx -(R/R_j)^3 (v_s R_j)/(3\beta c_s)$.

In Figure \ref{fig:shape}, we give an example of the shape of the 
bow shock shell, together with the observer-frame 
vector fields for the mean shell velocity
$(\bar v_R, \bar v_z')$ and the velocity of the newly swept-up shell material
$(u_R, u_z')$.  For this figure, we take the 
value of the ratio $ \beta c_s/v_s=0.5$; for the range of observed jet
velocities, the range of this ratio would be $\sim 0.2-0.6$.
In Figure \ref{fig:velplot}, we display the values of the various shell 
velocity components as seen from the observer frame, for the same model.

At large $R$, the component velocities of the newly-added material
approach $u_R\rightarrow \beta c_s R_j^2/R^2$ and 
$u_z'\rightarrow  (\beta c_s/v_s)^2 (R_j/R)^4 v_s$.  
That is, the transverse
velocity of newly-added material is the same as the existing mean
value of the transverse velocity, while the longitudinal velocity of
newly-added material is smaller than the mean longitudinal velocity by
the inverse of the large factor $(v_s R)^2/(\beta c_s R_j)^2$.  
The ratio of the
mean velocity components has a small, constant value $\bar v_R/ \bar
v_z'= \beta c_s /v_s$.
The newly-shocked material, on the other
hand, has an increasingly large ratio of transverse- to longitudinal-
velocity as the distance from the head of the jet increases,
$u_R/u_z'=\tan\theta\approx 1/(\pi/2-\theta)\rightarrow 
(v_s/\beta c_s) (R/R_j)^2$.

Near the head of the jet ($R/R_j\rightarrow 1$), the longitudinal 
speeds $u_z'$ and $\bar v_z'$ both approach the bow shock speed $v_s$.
The average transverse speed in the flow $\bar v_R$ approaches 
$\beta c_s\equiv v_0$, 
while the transverse speed of newly swept-up material $u_R$
reaches a local maximum value of $v_s/2$ near $R=R_j[1+ (\beta c_s/2 v_s)]$ 
and then declines to zero at $R=R_j$.  

The velocity $w_R(R;R_0)$, which would describe the transverse motion of
the material initially expelled from the WS in the absence of any
shear mixing with latterly-added fluid elements, remains close to both
$\bar v_R$ and $u_R$ far from the head of the shock (large $-z$).  
Its longitudinal counterpart, $w_z'(R;R_0)$,
however, remains much larger than both of the velocities 
$\bar v_z'$ and $u_z'$ that describe the mean longitudinal motion and the 
longitudinal motion of newly-added material.

In the shock frame, all of the transverse momentum in the shell is
provided by the initial impulse at the jet head (by assumption). The
transfer of portions of this transverse momentum to the newly-shocked
material is mediated by pressure, with the egalitarian result that the
transverse velocity of newly swept-up material is nearly the same as the
existing transverse velocity in the shell.  Longitudinal momentum, on
the other hand, is carried into the shell by every ambient mass
element that is swept up by the advancing shock.  Because the bow
shock is increasingly oblique at large distance from the jet head,
increasingly large portions of this longitudinal momentum can be
retained by the ambient material immediately after it enters the bow
shock.  As a consequence, the longitudinal flow velocity of the
newly-added material will be more negative (in the shock frame) than
the mean longitudinal flow velocity of the existing material in the
shell at the point of impact.  In the observer's frame, this
translates to a larger (positive) {\it mean} longitudinal velocity
$\bar v_z'$ than that of the newly-added material, $u_z'$.  The
newly-added material only speeds up in the forward direction to the
extent that mixing in the shear flow allows it to.  As we show from
the simulations in \cite{Lee2000b}, this occurs to a certain extent,
but in fact a significant level of velocity shear remains across the
thickness of the bow shock at any position.

\section{Kinematic diagnostics and macroscopic outflow properties}

>From the solutions obtained in the previous section, it is clear that 
the values at any position in the shell 
of the velocity transverse to the jet are 
quite insensitive to the degree of radial mixing of newly-added and
previously-existing shocked material.  That is, the values $\bar v_R$ and
$u_R$ describing the mean velocity and that of newly swept-up material 
are quite close to each other at any $R>>R_j$ (see e.g. Fig 
\ref{fig:velplot}a).  On the other hand, the values of the longitudinal 
velocity in the shell are fairly sensitive to the degree of mixing in the 
shell.  Since it is uncertain how thorough local mixing in fact will be
(this may also be affected by magnetic fields), 
kinematic diagnostic predictions based on this simple model are most
robustly applied to observational cases in which the jet lies close to 
the plane of the sky.  

We consider two diagnostics that yield qualitatively different
characteristic properties for wide-angle-wind-driven vs. jet-driven shells
\citep{Lee2000a,Lee2000b}. 
One of these diagnostics is the relation between the offset along the 
projected outflow axis from the jet head toward the stellar source
and the observed molecular shell velocity at that postion.
For plane-of-sky jets, the offset position is equal to $z$, and the 
line-of-sight velocity is $\pm v_R$. Thus, the position-velocity (``PV'')
relation is given (parametrically via the variable $R$) by equations
(\ref{eq:shapesol}) and (\ref{eq:meanvelR}) or (\ref{eq:newvelR}). 
At large distance $|z|$ from the jet head, the position-velocity relation
approaches $v_{obs}= [R_j/(3|z|)]^{2/3}(\beta c_s/v_s)^{1/3} v_s$; i.e. 
the velocity decreases as the inverse 2/3 power of the offset distance
from the head of the jet.  The expected characteristic feature in PV diagrams 
where a bow shock is present in a plane-of-sky jet is thus a symmetric 
(red/blue) pair of ``spurs'' aligned convex-inward along the $v=0$ axis.

The second diagnostic relation we consider is the distribution of 
mass with observed velocity.  For plane-of-sky jets, the observed 
velocity at any point in the shell is equal to $v_{obs}=v_R\cos\phi$,
where $\phi$ is the azimuthal angle in the shell.
We can therefore evaluate the distribution of mass with 
velocity at fixed $\phi$ as

\begin{equation}\label{eq-delM}
{\frac{\delta M}{\delta v_{obs}}}
=\frac{1}{\cos\phi}\frac{dM}{dv_R}\frac{\delta \phi}{2\pi}
=\frac{1}{\cos\phi}\frac{dM}{dR}\frac{dR}{dv_R}\frac{\delta \phi}{2\pi}
\end{equation}
where
\begin{equation}\label{eq-dMdR}
\frac{dM}{dR}= \frac{\dot{M}}{v_R}=\frac{\dot{P}_{oR}}{v_R^2}
=\pi \rho R_j^2 v_s \beta c_s\left({\cos\phi\over v_{obs}}\right)^2
\end{equation}
and
\begin{equation}\label{eq-dRdv}
\frac{dR}{dv_R}=-\frac{\dot{P}_{oR}}{2 \pi v_s \rho R v_R^2}
=-{1\over 2}R_j(\beta c_s)^{1/2}  \left({\cos\phi\over v_{obs}}\right)^{3/2}.
\end{equation}
For any value of $v_{obs}>0$, $\phi$ must fall in the range 
$|\phi|\le \cos^{-1}[v_{obs}/(\beta c_s)]$.
Substituting equations (\ref{eq-dMdR})-(\ref{eq-dRdv}) in equation 
(\ref{eq-delM})
and integrating over $\phi$, we find for $v_{obs}<<\beta c_s$ (so that 
$|\phi|\le \pi/2$) 
\begin{equation}
m(v_{obs})\equiv {dM\over dv_{obs}}=
-0.3594{\rho R_j^3 v_s (\beta c_s)^{3/2}\over v_{obs}^{7/2}}.
\end{equation}
When $v_{obs}\rightarrow\beta c_s$, the profile is cut off with
$m(v_{obs})  \propto [1-(v_{obs}/\beta c_s)]^{1/2} v_{obs}^{-7/2}$.
For plane-of-sky sources, the red and blue sides of the profile are
symmetric.

We compare these predicted position-velocity and mass-velocity
diagnostic relations to the results of simulations and molecular line
observations in \cite{Lee2000b}.

In addition to these detailed ``microscopic'' diagnostics, which are
valuable for direct  comparisons with high-resolution observations, 
it is useful to summarize the dependence of the macroscopic outflow 
properties on the basic physical parameters involved.  These 
macroscopic characteristics include the width-to-length ratio of the 
shell, its total mass, and the ratio of components of total 
shell momentum perpendicular and parallel to the jet axis.  

The width/length of the jet shell at distance $|z|$ from the head of the 
jet is given by $2R/|z|$, which from equation (\ref{eq:shapesol}) at large
distance equals $2(3\beta c_s/v_s)^{1/3} |R_j/z|^{2/3}$, or, using physical
units and setting $|z|=v_s t$,
\begin{equation}\label{eq:width/length}
{width\over length}= 0.012 \left({t\over 10^4\yr }\right)^{-2/3} 
\left({\beta c_s\over 32 \kms}\right)^{1/3}
\left({R_j\over 100 \AU}\right)^{2/3}  \left({v_s\over 100\kms }\right)^{-1}. 
\end{equation}
Thus, except at the very earliest times, the outflow shell associated 
with a ``pure jet'' bow shock would show very extreme collimation (more 
than 100-to-1).  The width/length ratio of the shell is greater than the
width/length ratio of the jet itself by a factor 
$[3\beta c_s |z|/(v_s R_j)]^{1/3}$, of order 3-10 for typical parameters.
In Figure (\ref{fig:multscale}), we display the shape of the outflow shell 
at three different scales, from $12 R_j - 300 R_j$ in length, correponding
to lengths from a few thousand AU to a few tenths of a parsec.  

The total mass in the shell, obtained by integrating 
$dM/dR= \dot M/v_R= \pi \rho v_s R^4/(R_j^2 \beta c_s)$ over $R$,
can be written in terms of $t=|z|_{max}/v_s$ as 
\begin{equation}
M_{shell}(t)={3^{5/3}\pi\over 5} \left({\beta c_s } \right)^{2/3} v_s 
\rho R_j^{4/3} t^{5/3}.
\end{equation}
This shell mass can be compared to the mass in the jet itself,
$M_j(t) = \rho_j v_j \pi R_j^2 t$;  the ratio is given by 
$M_{shell}/M_{jet} = 3^{5/3} 5^{-1} (\beta c_s t/R_j)^{2/3} 
[\eta + \eta^{1/2}]^{-1}$, which is $\sim 100$  after $\sim 10^5$ years
for typical values of the parameters.

The ratio of total transverse to total longitudinal (observer frame) 
momentum $P_R/P'_z$ in the shell is
the same as the ratio $\bar v_R/\bar v_z'=\beta c_s/v_s$ given from equations 
(\ref{eq:meanvelz}) and (\ref{eq:meanvelR}), which is independent of 
position. Since this ratio is less than one,
the shell will {\it in bulk} have more forward-directed 
than sideways-directed motion, although {\it away from the jet head} the 
transverse motion exceeds longitudinal motion (since mixing is very incomplete
in the shell).  Because the ``bulk'' and spatially-resolved kinematics of
jet-driven shells are so different, it is crucial to obtain 
high-resolution observations in order to make discriminating comparisons with
theoretical models.

\section{Summary and discussion}

In this paper, we have constructed an analytic dynamical model for the
shape and kinematics of the bow shock shell created when a
protostellar jet impacts the surrounding (undisturbed) interstellar
medium.  Morphologically, the shell consists of two parts: the
``working surface,'' a hockey-puck-shaped region of radius $\sim R_j$
where the jet collides directly with the ambient medium, and
surrounding ``wings'' at $R>R_j$ which separate the low-density
``cocoon'' of shocked jet gas from the undisturbed ambient medium (see
Fig. 1).  The shell is composed of ambient material swept up by the
advancing bow shock, with densities and pressures high in working
surface and lower in the wings.  We analyze the flow in the wings in
detail; the working surface region is, in our model, analyzed solely
to estimate the mass and momentum flow it ejects into the surrounding
medium.

The chief simplifying assumption we invoke in our analysis is that
thermal pressure forces play an important role only in the working
surface at the head of the jet, but not in the wings of the bow shock.
An initial transverse momentum impulse (perpendicular to the jet, and
locally tangential to the shell's surface) and mass input are applied
to the shell flow in the wings at $R=R_j$.  Subsequent to this
impulse, the shell flow expands away from the axis and sweeps up
ambient material in a ballistic fashion -- i.e.  conserving the total
mass and momenta of the shell + swept-up ambient material.
Physically, it is the radial pressure gradients within the working
surface that impart the initial transverse impulse to the shell in the
wings.  The initial mass flux to the wings is provided by ambient gas
which has entered the working surface through its outer face, then been 
radially redirected and expelled from the sides of the working surface.  The
ballistic shell's shape and the velocities in the wings depend on the ambient
density, the shock speed $v_s$, and on the initial values of mass and
momentum flows emerging perpendicularly from the working surface (but
not on its detailed internal dynamics).

\begin{figure}
\center
\epsscale{1.}
\plotone{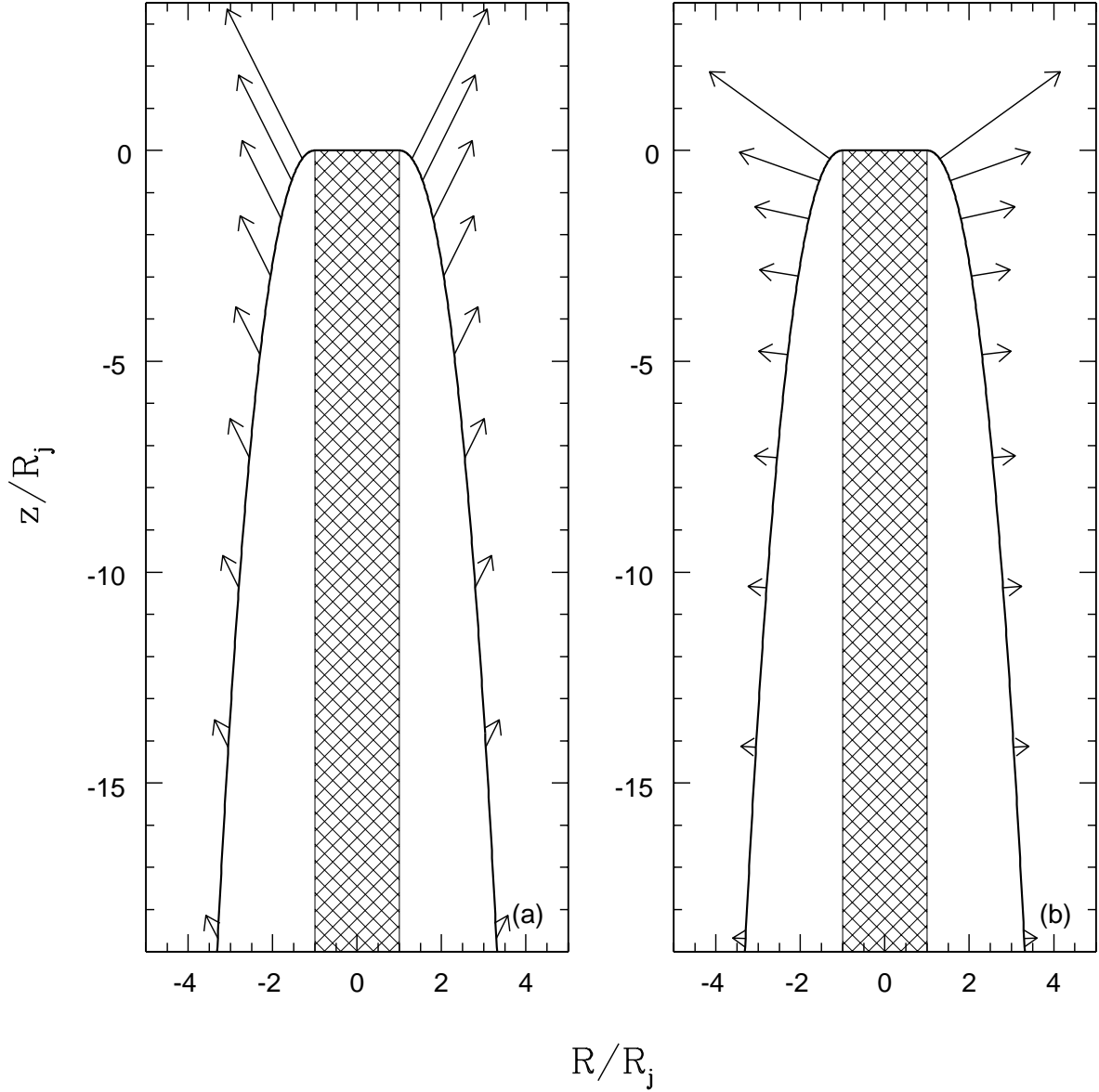}
\caption{Shell shape and velocities in the wings for $\beta c_s/v_s=0.5$. 
Panel (a) shows the mean shell velocities $\bar {\bf v}'$, and panel (b)
shows the velocities of newly swept-up material ${\bf u}'$, in the observer 
frame.   The jet beam is shown as a crosshatched region.
\label{fig:shape}}
\end{figure}

Because the shocked ambient gas in the working surface cools rapidly
to $\sim 10^4$K, with corresponding sound speed $c_s\sim 8\kms$, the
transverse velocity $v_R\sim c_s$ of the gas input to the shell from
the working surface will be small compared to its longitudinal
(observer reference frame) velocity $v_z=v_s$ along the jet axis.
Memory of this ratio is preserved as the mean ratio $\beta c_s /v_s$
(with $\beta$ an order-unity constant) of transverse/longitudinal
momenta in the wings of the shell (see Fig. (\ref{fig:shape}a)).  
This strongly forward-directed
mean thrust is responsible for the high degree of elongation that
develops in the bow shock shell (see Fig \ref{fig:multscale}).  
The ultimate cause of
the extreme aspect ratio in these bow shocks can thus be attributed to
the effectiveness of interstellar cooling: the sound speed in the
shocked ambient gas in the jet head does not remain near its immediate
post-shock value, $\sim v_s$, but instead drops to a much lower level,
with the consequence that the transverse pressure-gradient thrust is
relatively weak.  Only when cooling is minimal, as occurs for radio jets
plowing into the intergalactic medium, the transverse thrust (which
also includes effects of pressure forces over the body of the shell
from hot shocked jet gas in the cocoon) can be comparable to the
longitudinal thrust, with the consequence that the lobes created are
much less elongated.

\begin{figure}
\center
\epsscale{1.}
\plotone{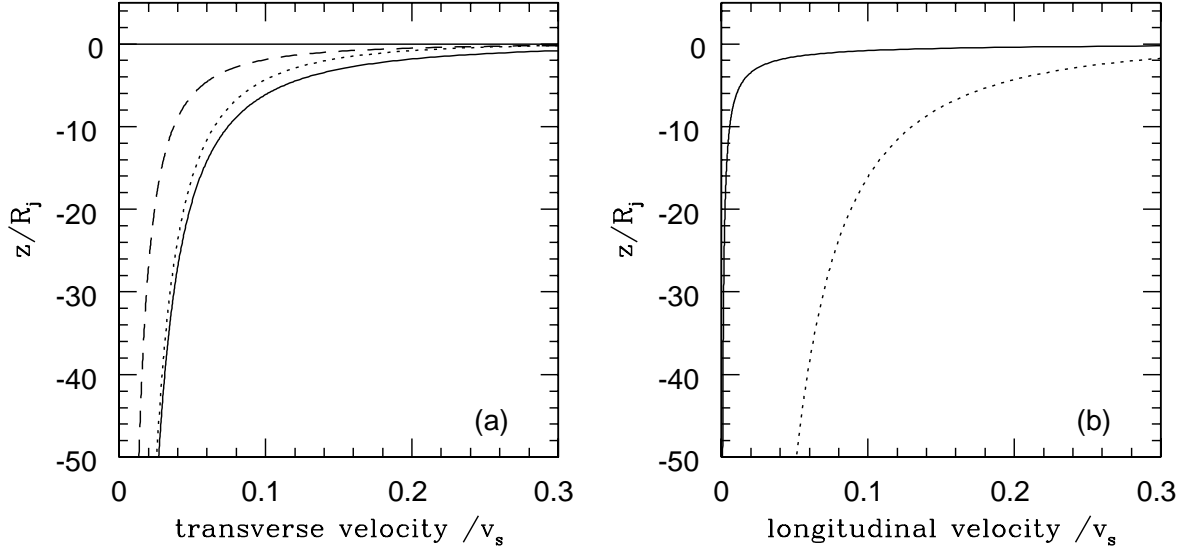}
\caption{Shell velocities in observer frame, for $\beta c_s/v_s=0.5$ 
model. Frame (a) shows $u_R$, $\bar v_R$, and $w_R(R;R_0)$ with solid, 
dotted, and dashed lines, respectively.  Frame (b) shows $u_z'$ and 
$\bar v_z'$ with solid and dotted lines, respectively 
($w_z'(R;R_0)$ is everywhere too large to appear in the frame).
>From equations (\ref{eq:meanvelR}) and (\ref{eq:newvelR}), note that 
$\bar v_R$ and $u_R$ are approximately proportional to $\beta c_s$ at large 
distance from the jet head.
\label{fig:velplot}}
\end{figure}

\begin{figure}
\center
\epsscale{1.}
\plotone{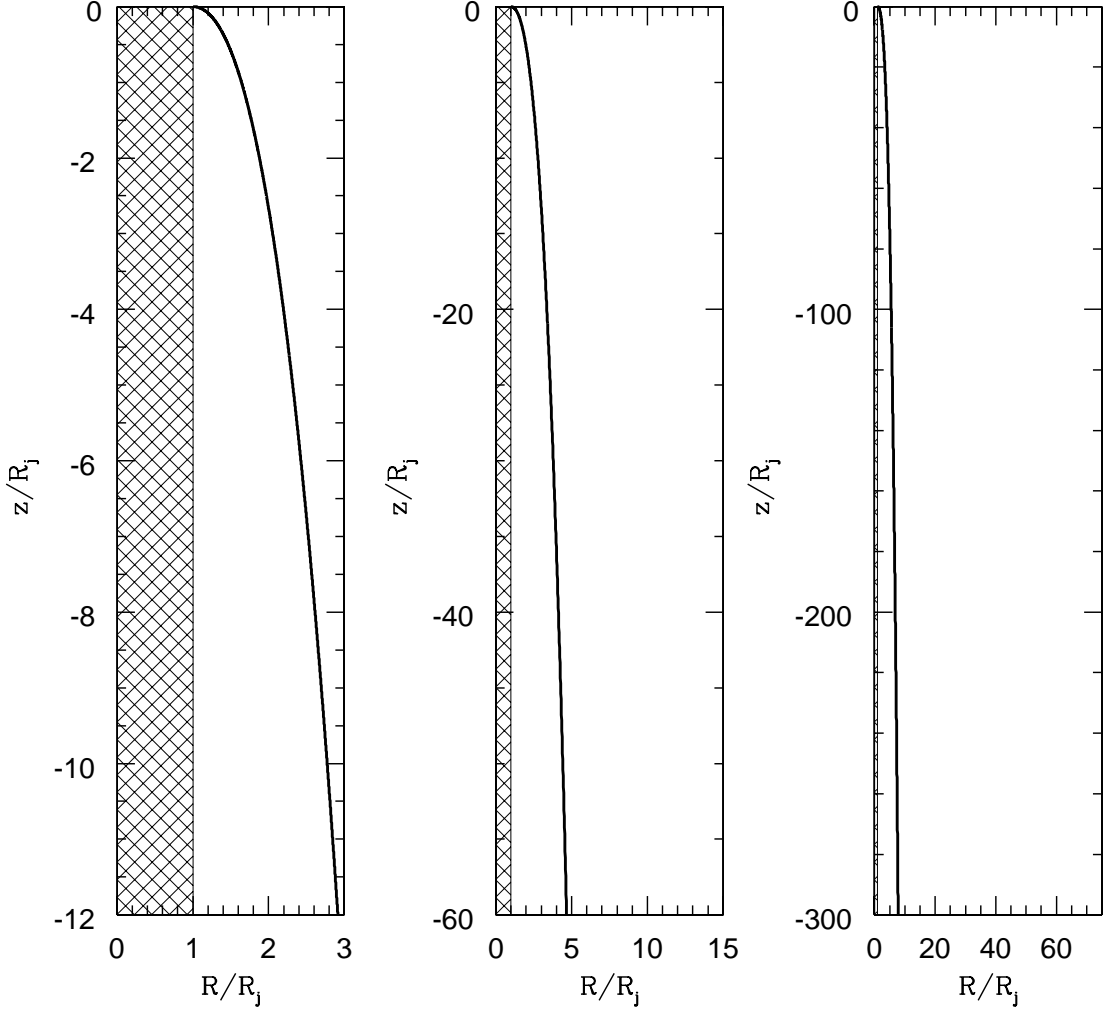}
\caption{Outflow shell shape on multiple scales, for model with 
$\beta c_s/v_s=0.5$.  In each panel, the central crosshatched region 
demarks (half of) the jet beam;
the heavy solid line shows the locus of the shell.  Left, center, and right
panels have vertical scales $\sim 0.01$, $0.06$, and $0.3$ pc, 
respectively, for $R_j=200$AU.
\label{fig:multscale}}
\end{figure}

Our analysis and basic results on bow shock shell shapes and
kinematics are presented in \S 2.  We give the shape of the bow shock
shell in equation (\ref{eq:shapesol}), the crossection-averaged mean
shell velocity components $\bar v_z'$ (observer frame) and $\bar v_R$
in equations (\ref{eq:meanvelz})-(\ref{eq:meanvelR}), and the velocity
components of the outer surface layer of the shell $u_z'$ and $u_R$ in
equations (\ref{eq:newvelz})-(\ref{eq:newvelR}).  Because of
incomplete mixing between ``newly-added'' and ``existing'' shell
material at any point, there may be shear in the component of velocity
parallel to the shell surface.  The values of the mean velocity and
the velocity of the surface layer thus bracket the range of velocities
expected in the shell at any point $z$ (see Figs. \ref{fig:shape},
\ref{fig:velplot}).  We further use our analytic results of \S2 to
present, in \S3, the predicted position-velocity and mass-velocity
relationships for the bow-shock shells produced by plane-of-sky jets,
and to summarize the bulk properties of outflow shells generated via
leading jet bow shocks as analyzed in this paper.

In the companion paper to this one \citep{Lee2000b}, we show that the
analytic model developed here provides excellent agreement with the
shell shapes found in numerical simulations of jet-driven bow shocks.
We also find that the transverse velocities derived in the analytic
model closely agree with those computed in the simulations.
We find that the bow shock material predominantly has observer-frame
axial velocity closer to $u_z'$ (the value associated with
``newly swept-up'' gas), although the value $\bar v_z'$ (the predicted
crossection-averaged axial speed) provides a good fit to the upper
envelope of the distribution of longitudinal shell velocities in the
simulations.  We also find that the predicted behavior of the
mass-velocity relation $m(v_{obs})\propto v_{obs}^{-7/2}$ is in good
agreement with the results of simulations, for plane-of-sky jet axes.

The excellent agreement between our analytic results and specific
numerical models gives us confidence in drawing on the former for more
general predictions about the structure and evolution of jet-driven
bow shocks.  The overarching goal, of course, is to assess whether
observed molecular outflows are likely produced principally by the
jet-driven-bow-shock mechanism.  Although we do think that many
prominent (especially high-velocity) {\it features} in molecular flows
are chiefly the result of jet bow shocks (see below), we 
conclude from this work that 
assembly of ``classical'' large-scale bipolar lobes must incorporate other
ingredients.  The basic physical reason for this conclusion is that,
under strongly-cooling interstellar conditions, pressure gradient
forces in the shocked ambient (and jet) gas are insufficient to create
transverse momentum fluxes comparable to the longitudinal fluxes
inherited from the jet beam itself, with the consequence that the
lobes delimited by the bow shock shells will be highly elongated.
Equation (\ref{eq:width/length}) embodies this conclusion
quantitatively: the shell width/length 
varies as the $1/3$ power of the (small) ratio $\beta c_s/v_s$ of 
transverse/longitudinal momentum imparted to the shell by the jet, and 
also decreases as the $-2/3$ power of time.\footnote{Our assumption of 
a cylindrical rather than conical jet beam enhances the time-dependent 
effects on the width/length ratio, but is in fact consistent with the 
asymptotic density distribution in both collimated and force-free MHD winds.}
Within a thousand years, the width/length ratio of the bow shock drops 
below the 1:10 ratio typically associated with bipolar molecular flows.
This progressive elongation is strikingly illustrated in Figure 
(\ref{fig:multscale}), which portrays the outflow shape at multiple scales;
these correspond to successive factors of five increase in the age of the 
jet.

Our conclusion that a single leading bow shock is unable to produce
large-scale molecular outflows is consistent with the findings of
previous authors from both semianalytic \citep{Masson1993} and
numerical studies (see references in \cite{Lee2000b}).  Advocates of
``pure jet'' models argue that internal bow shocks or jet wandering
can help widen the bow-shock shells.  Because internal bow shocks
produce much less transverse momentum flux than leading bow
shocks (by a factor $\sim \Delta v/v_j$, where $\Delta v$ is the
velocity variation in the jet beam that leads to the internal shock),
however, we do not expect their contribution to provide a major
boost to the transverse shell expansion, and this is indeed what our
numerical simulations show (\cite{Lee2000b}; see also
\cite{Stone1993a, Stone1993b}).  Jet precession, if sufficiently rapid
to produce a smooth shell, is indistinguishable from an intrinsically
wide-angle wind distribution.  Thus, we contend that the relatively
weak collimation of ``classical'' bipolar outflows cannot be
reconciled with the idea that an optical jet represents the whole
angular extent of the primary wind from a protostar; we infer that at
least {\it some} wide-angle wind must be present.

Although jet bow shocks cannot explain everything, 
they can explain molecular shell features associated
with ``spurs'' seen in position-velocity diagrams of high spatial
resolution CO maps \citep{Lee2000a}.  The analysis of this paper
focuses on leading bow shocks, because the initial conditions are very
well defined.  Internal bow shocks can, however, be analyzed in a
similar fashion.  In a pure jet model, the cocoon gas enveloping
the jet beam (but still interior to the leading bow shock) will be at low 
density
$\rho_e$ compared to the ambient ISM (it chiefly consists of shocked
jet gas expanded in volume), and will have significant forward motion
$v_e$ ($\sim v_{s0}$, where $v_{s0}$ is the velocity of the next shock 
downstream).
In a model where the jet is a dense core within a wide-angle wind,
again the surrounding envelope gas would have low density (by assumption) and
large forward velocity ($v_e\le v_j$).  The internal shock speed
$v_{si}=(v_1+v_2)/2$ is the mean value of the upstream ($v_2$) and
downstream ($v_1$) jet speeds.  Equations
(\ref{eq:jetM})-(\ref{eq:uR}) would all carry through, with the
replacement of $v_s\rightarrow v_{si} - v_e $, and $\rho\rightarrow \rho_e$. 
The input mass
and transverse momentum fluxes take the same form as before, except
with $\rho\rightarrow \rho_j$ and $v_s\rightarrow (v_2-v_1)/2=\Delta
v/2$.  The shell shape still has a $|z|\propto R^3$ behavior at large
distance (with appropriately modified coefficient), and the
position-velocity relation still approches the form $v_R\propto
|z|^{-2/3}$ (again with modified coefficient).
\footnote{Explicitly, the shell shape takes the form of equation 
(\ref{eq:shapesol}), except with $v_s$ replaced by 
$2 (\rho_e/\rho_j)(v_{si}-v_e)^2/\Delta v$. The 
crossection-averaged mean shell transverse velocity $\bar v_R$ becomes
\begin{equation}
\beta c_s \left[\left({2\rho_e\over \rho_j} \right) \left({v_{si}-v_e \over \Delta v} \right) \left({R^2\over R_j^2} -1\right) +1 \right]^{-1}, 
\end{equation}
which at large distance approaches the form of equation
(\ref{eq:meanvelR}) divided by $2 (\rho_e/\rho_j)(v_{si}-v_e)/\Delta
v$.  The position-velocity relation therefore approaches 
$v_R= (v_{si}-v_e) {\cal A}^{-1/3} (3|z|/R_j)^{-2/3}$, where
${\cal A}\equiv  2 (\rho_e/\rho_j)(v_{si}-v_e)^2/(\beta c_s\Delta v )$.  
Larger shock jump velocities and smaller envelope densities tend to 
increase the value of the transverse shell velocity at a given offset 
position $z$.}  In \cite{Lee2000b}, we demonstrate that the latter functional
form indeed yields a good fit to the ``spur'' structures in the
observed outflow HH 212.

\clearpage

\begin{thebibliography}{}

\bibitem[Bachiller \& Tafalla(1999)]{Bachiller1999} Bachiller, R., \& 
Tafalla, M. 1999, in The Origin of Stars and Planetary Systems, eds.
C.J. Lada \& N.D. Kylafis (Dordrecht:Kluwer), p. 227 

\bibitem[Biro \& Raga(1994)]{Biro1994} Biro, S., \& Raga, A.C. 1994, 
\apj, 434, 221

\bibitem[Chernin et al(1994)]{Chernin1994} Chernin, L., Masson, C., Gouveia
Dal Pino, E.M., \& Benz, W. 1994, \apj 426, 204


\bibitem[Downes \& Ray(1999)]{Downes1999} Downes, T. P., \&  Ray, T. P. 
\aap 345, 977

\bibitem[Falle \& Raga(1993)]{Falle1993} Falle, S.A.E.G., \& Raga, A.C. 
1993, \mnras, 261, 573

\bibitem[Kim \& Ostriker(2000)]{Kim2000} Kim, W.-T., \& Ostriker, E.C.
2000, \apj, in press

\bibitem[K\"onigl \& Pudritz(2000)]{Konigl2000} K\"onigl, A., \& 
Pudritz, R.E. 2000, in Protostars
  and Planets IV, ed. V. Mannings, A. P. Boss \& S. S. Russell
  (Tucson: University of Arizona Press), p. 759


\bibitem[Lee et al.(2000a)]{Lee2000a} Lee,  C.-F., 
Mundy, L. G., Reipurth, B. 
Ostriker, E.C.,  \& Stone, J.M.  2000a, \apj, submitted 

\bibitem[Lee et al.(2000b)]{Lee2000b} Lee, C.-F., Stone, J.M.,
Ostriker, E.C., \& Mundy, L. G. 2000b, \apj, submitted 

\bibitem[Li \& Shu(1996)]{Li1996} Li, Z. -Y.  \& Shu, F. H. 
  1996, \apj, 472, 211 

\bibitem[Masson \& Chernin(1993)]{Masson1993} Masson, C.R., \& Chernin, L.M.
1993, \apj, 414, 230

\bibitem[Matzner \& McKee(1999)]{Matzner1999} Matzner, C. D. \& 
  McKee, C. F. 1999, \apjl, 526, L109 

\bibitem[Moriarty-Schieven et al(1987)]{Moriarty1987}  Moiarty-Schieven, 
Moriarty-Schieven, G. H., Snell, R. L., Strom, S. E., Schloerb, F. P.,
Strom, K. M., \& Grasdalen, G. L. 1987, \apj 319, 742

\bibitem[Nagar et al(1997)]{Nagar1997} Nagar, N.M.,
 Vogel, S.N., Stone, J.M., \& Ostriker, E.C. 1997, \apj, 482, L195

\bibitem[Ostriker(1997)]{Ostriker1997} Ostriker, E. C. 1997, \apj, 486, 291

\bibitem[Ostriker(1998)]{Ostriker1998} Ostriker, E.C. 1998, 
in Accretion Processes in Astrophysical Systems, eds. S.Holt \& T. Kallman
(Woodbury NY:AIP Press), p. 484


\bibitem[Raga \& Cabrit(1993)]{Raga1993} Raga, A., \& Cabrit, S. 1993,
\aap, 278, 267

\bibitem[Richer et al(2000)]{Richer2000} Richer, J.S., Shepherd, D.S., 
Cabrit, S., Bachiller, R., \& Churchwell, E. 2000,
in Protostars
  and Planets IV, ed. V. Mannings, A. P. Boss \& S. S. Russell
  (Tucson: University of Arizona Press), p. 867

\bibitem[Shu et al.(1991)]{Shu1991} Shu, F. H., 
  Ruden, S. P., Lada, C. J. \& Lizano, S.  1991, \apjl, 370, L31 

\bibitem[Shu et al.(1995)]{Shu1995} Shu, F. 
  H., Najita, J. , Ostriker, E. C. \& Shang, H.  1995, \apjl, 455, L155  

\bibitem[Shu et al.(2000)]{Shu2000} Shu, F.H.,  Najita, J., Shang, H., 
  \& Li, Z. -Y. 2000, in Protostars
  and Planets IV, ed. V. Mannings, A. P. Boss \& S. S. Russell
  (Tucson: University of Arizona Press), p. 789

\bibitem[Smith, Suttner, \& Yorke(1997)]{Smith1997} Smith, M.D., Suttner, G.,
\& Yorke, H.W. 1997, 323, 223

\bibitem[Stone \& Norman(1993a)]{Stone1993a} Stone, J.M., \& Norman, M.L.
1993, \apj 413, 198

\bibitem[Stone \& Norman(1993b)]{Stone1993b} Stone, J.M., \& Norman, M.L.
1993, \apj 413, 210

\bibitem[Suttner et al(1997)]{Suttner1997} Suttner, G., Smith, M.D., Yorke,
H.W., \& Zinnecker, H. 1997, \aap 318, 595

\bibitem[Wilkin(1996)]{Wilkin1996} Wilkin, F.P. 1996, \apj, 459, L31

\end{thebibliography}
\end{document}